\documentstyle[12pt]{article}
\input epsf

\begin{document}

\def\be{\begin{equation}}
\def\ee{\end{equation}}
\def\bea{\begin{eqnarray}}
\def\eea{\end{eqnarray}}
\def\N{{\cal N}}
\def\S{{\cal S}}
\def\a{{\alpha}}
\def\b{{\beta}}
\def\d{{\delta}}
\def\G{{\Gamma}}
\def\t{{\tau}}
\def\cL{{\cal L}}
\def\L{{\Lambda}}
\def\cF{{\cal F}}

\newpage
\bigskip
\hskip 3.7in\vbox{\baselineskip12pt
\hbox{NSF-ITP-01-01}}

\bigskip\bigskip

\centerline{\large \bf  Comments on fractional instantons 
in $\N=2$ gauge theories}

\bigskip\bigskip

\centerline{{\bf
Alex Buchel\footnote{buchel@itp.ucsb.edu}}}

\bigskip
\centerline{Institute for Theoretical Physics}
\centerline{University of California}
\centerline{Santa Barbara, CA\ \ 93106-4030, U.S.A.}

\begin{abstract}
\baselineskip=16pt
$\N=1^{\star}$ gauge theories are believed to have fractional 
instanton contributions in the confining vacua. 
D3 brane probe computations in gravitation dual of large-N 
$\N=2^{\star}$ gauge theories point to the absence of such contributions 
in the low energy  gauge dynamics. We study fractional instantons 
in $\N=2$ $SU(2)$ Yang-Mills theory from the field theoretical 
perspective. We present new solutions to the Seiberg-Witten 
$SU(2)$ monodromy problem with the same perturbative 
asymptotic, a massless monopole and a dyon singularity on the 
moduli space, and fractional instanton corrections to 
$\N=2$ prepotential in the semi-classical region of the 
moduli space. We show that fractional instantons lead 
to infinite monopole (dyon) condensate in mass deformed $\N=2$
gauge theories.

\end{abstract}
\newpage
\setcounter{footnote}{0}

\section{Introduction}

Study of non-perturbative effects in 4D gauge theories is an important 
research direction. Unlike ordinary gauge theories where often the 
only way of understanding strongly coupled dynamics is to 
do numerical simulation, a large class of phenomena in supersymmetric
gauge theories can be understood analytically and exactly.   
The latter is due to strong restrictions implied by the supersymmetry 
on the possible structure of perturbative and non-perturbative 
effects. Typically, the larger the supersymmetry in the theory,
the more constraint is its dynamics. Perturbatively, in gauge theories 
with  $\N=1$ SUSY the superpotential is not renormalized 
 \cite{gsr}, the beta function 
in $\N=2$ theories is not modified beyond one-loop order 
\cite{pertprep}, and the $\N=4$ theory is finite \cite{n4}. 
There are no non-perturbative corrections to $\N=4$ beta function, 
while perturbative beta function of gauge theories with 8 (or less) 
supercharges is corrected by instantons \cite{s88}. 

An interesting question to ask is whether 
non-perturbative corrections in supersymmetric gauge 
theories are due only to instantons. There is strong evidence,
both from the field theory perspective \cite{kss9706}, and 
within the framework of the  D-brane engineering 
of gauge theories \cite{b9803}, that fractional instantons,
carrying $1/N$ units of instanton charge,  are responsible 
for gaugino condensation and the mass gap in low energy
$\N=1$ $SU(N)$ supersymmetric Yang-Mills theory. 
Furthermore, the superpotential of the mass deformed 
$\N=4$ $SU(N)$ YM-theory (also known as the $\N=1^{\star}$ theory)
was shown in the confining vacua to have fractional
instanton expansion \cite{dk0001}. The latter result
was confirmed in the supergravity dual of $\N=1^{\star}$
gauge theory constructed by Polchinski and Strassler
\cite{ps0003}. On the contrary, D-brane construction of 
$\N=2$ gauge theories \cite{w97,b9803} and the analysis 
of a D3-probe dynamics in gravitational dual of $\N=2^{\star}$ 
gauge theories \cite{bpp0008,ejp} suggests that 
fractional instantons do not play role in the low 
energy dynamics of these theories.

The purpose of this paper is to study the consequences 
of presence of fractional instantons in 4D $\N=2$ gauge theories
from the field theoretical perspective. As it is well 
known, in the standard solution for the low-energy effective 
action of $\N=2$ $SU(2)$ SYM theory \cite{SW}, only 
integer instantons contribute to the prepotential.
This is actually an
input to the solution, rather than its prediction.
In fact, we construct infinitely many new solutions to the Seiberg-Witten
$SU(2)$ monodromy problem with the same (perturbative)
weak coupling asymptotic, and a pair of additional singularities 
on the moduli space where a monopole and a  dyon, with the 
same charges as in \cite{SW}, respectively becomes massless.
These new solutions differ from the original one in that 
the gauge coupling of the low-energy effective action 
receives non-perturbative corrections of $1/2$ unit the 
instanton charge. Though mathematically acceptable, 
all the new solutions we find fail physically. Specifically,
in \cite{SW}, partial supersymmetry breaking by a 
soft mass term to the chiral superfield in the $\N=2$ 
vector multiplet lifted all vacua of the moduli space, 
except for those with massless monopole and dyon. 
One is left then with two $\N=1$ vacua as predicted 
in \cite{w82}. In our solutions, all the Coulomb 
branch is lifted under the soft SUSY breaking: 
the monopole (dyon) condensate is infinite 
at the moduli space singularities. 

The paper is organized as follows. In the next section 
we discuss the would-be signature of fractional instantons in supergravity
dual of large-N gauge theories with 8 supercharges. 
Field-theoretical analysis of $\N=2$ $SU(2)$ YM theory 
with fractional instanton contributions to the prepotential
is presented in section 3. We conclude in section 4.

\section{Fractional instantons in $\N=2$ gauge theories 
in the framework of Maldacena duality}

AdS/CFT duality of Maldacena \cite{juan} relates strongly coupled 
superconformal gauge theories to supergravity backgrounds.
In \cite{juan} it was shown that
four dimensional $\N=4$ $SU(N)$ Yang-Mills 
theory, at large values of 't Hooft coupling and $N$, 
has a weakly coupled 
description as type IIB string theory compactified on $AdS_5\times 
S^5$ with $N$ units of the five-form flux through the $S^5$.
$AdS_5\times S^5$ gravitational background represents a near horizon 
geometry of coincident $N$ D3 branes, and is dual to the 
origin of the Coulomb moduli space of the gauge theory.
The classical $3(N-1)$ complex dimensional moduli space 
of the gauge theory is not corrected quantum mechanically, 
and has a very simple interpretation in supergravity as a moduli
space of multi-centered solutions with D3-brane charge \cite{juan,klt}.
The fact that there are no quantum corrections to the moduli
space  relates to the possibility of moving
without obstruction a D3-brane probe in $AdS_5\times S^5$ background.   

The situation is different in the case of gravitation dual 
of gauge theories with reduced supersymmetry. Gauge theories 
with 8 supercharges has quantum moduli space which is however 
different from the classical one. A typical example is the 
classical vacuum of the four dimensional $\N=2$ $SU(2)$ YM
theory with unbroken gauge symmetry which does not survive the 
quantization \cite{SW}. This has profound implications for the 
probe dynamics in dual gravitational backgrounds. To be more specific 
we concentrate now on the probe computation in the gravitational 
background dual to the mass deformed $\N=4\to \N=2$ $SU(N)$ 
Yang-Mills theory, also known as $\N=2^{\star}$ gauge theory.
The corresponding supergravity background, which we refer to as 
PW, was constructed 
in \cite{pw} and the D-brane probe dynamics was discussed in 
\cite{bpp0008,ejp}. 
In the language of four-dimensional $\N=1$ supersymmetry, the mass
deformed $\N=4$ $SU(N)$ Yang-Mills theory consists of a vector
multiplet $V$, an adjoint chiral superfield $\Phi$ related by $\N=2$
supersymmetry to the gauge field, and two additional adjoint chiral
multiplets $Q$ and $\tilde{Q}$ which form the $\N=2$ hypermultiplet.  In
addition to the usual gauge-invariant kinetic terms for these fields,
the theory has additional interactions and hypermultiplet mass term
summarized in the superpotential\footnote{The classical K\"{a}hler
potential is normalized $(2/g_{YM}^2){\rm tr}[\bar{\Phi}\Phi+
\bar{Q}Q+\bar{\tilde{Q}}\tilde{Q}]$.}
\be
W={2\sqrt{2}\over g_{YM}^2}{\rm tr}([Q,\tilde{Q}]\Phi)
+{m\over g_{YM}^2}({\rm tr} Q^2+{\rm tr}{\tilde Q}^2)\,.
\label{sp}
\ee
The theory has a classical moduli space of Coulomb vacua parameterized by
expectation values of the adjoint scalar
\be
\Phi={\rm diag} (a_1,a_2,\cdots,a_N)\,,\quad \sum_i a_i=0\,,
\label{adsc}
\ee
in the Cartan subalgebra of the gauge group.  For generic values of
the moduli $a_i$ the gauge symmetry is broken to that of the Cartan
subalgebra $U(1)^{N-1}$, up to the permutation of individual $U(1)$
factors. 
Classically, when two or more moduli $a_i$ coincide,  the gauge symmetry 
is appropriately enhanced.  
The moduli space of a D3-brane probing the PW 
supergravity background is dual to the projection of the 
Coulomb branch vacua of $SU(N+1)\to U(1)\times U(1)^{N-1}$ to that of the
probe $U(1)$. If $u$ is the modulus of the $U(1)$ representing the
probe, the classical parametrization of the full moduli space
(\ref{adsc}) is given by
\be
\Phi={\rm diag} (u,a_1-u/N,a_2-u/N,\cdots,a_N-u/N)\,,\quad
\sum_i a_i=0 \,.
\label{adscu}
\ee
Up to coordinate change, $u$ identifies the position of the 
probe brane in the supergravity background. Classically, 
all values of $u$ are allowed. As we already mentioned, 
this is true quantumly for the corresponding modulus of  
$\N=4$ YM theory, resulting in the fact that a D3 probe 
can be moved freely in $AdS_5\times S^5$ background. 
In $\N=2^{\star}$ gauge theory, the classical $U(1)$ probe modulus 
$u$ receives quantum corrections \cite{dw}. Here, there are no 
perturbative corrections, but there are instanton 
corrections which become increasingly important as $u(1+1/N)\to a_i$
in (\ref{adscu})\footnote{More precisely,  in 
the large $N$ limit, instanton corrections 
become important as $|u-a_i|\ll |u|/N$ \cite{ds,bpp0008}.}. As in the case 
of $SU(2)$ YM theory, the vacua with classical nonabelian gauge symmetry,
$u(1+1/N)=a_i$, are not preserved by quantum corrections. 
As the result, one should expect boundaries 
in the D3-probe moduli space of the $\N=2^{\star}$ gravitational dual.  
The boundary of the D3-probe (or more generally Dp-probe) 
moduli space in gravitational dual of gauge theories 
with 8 supercharges is nothing but the enhancon of \cite{jpp9911}.
We would like to emphasize that this boundary is infinitely sharp 
in the $N\to \infty$ limit {\it only} if there are no fractional 
instanton corrections in theories with 8 supercharges.
This is precisely what was found in \cite{jpp9911,pw,bpp0008,ejp}.
The metric on the D3-probe moduli space is related by $\N=2$ supersymmetry 
to the imaginary part of the complexified probe $U(1)$ gauge 
coupling $\tau={\theta\over 2\pi}+i{4\pi\over g_{YM}^2}$ . This 
coupling receives both perturbative and 
nonperturbative corrections.  The perturbative corrections are one-loop
exact \cite{pertprep}. From the field theory perspective,
it was argued in \cite{bpp0008} that nonperturbative corrections  
due to instantons do not survive 't Hooft limit, thus the metric 
on the D3 probe moduli space is one-loop exact 
even  non-perturbatively. This result has been confirmed explicitly
in \cite{bpp0008} by comparing the one-loop $\tau$ computation in the 
$\N=2^{\star}$ gauge theory with induced metric on the D3 probe
moduli space in the PW geometry. Nonperturbative corrections 
carrying $p/N$  unit the instanton charge are 
suppressed as  $\exp(-8\pi^2 p/(N g^2_{YM}))$, 
and thus would contribute in the 't Hooft limit for any finite $p$,
invalidating one-loop gauge theory/supergravity agreement of \cite{bpp0008}.

\section{Fractional instantons in $\N=2$ $SU(2)$ Yang-Mills theory}

In this section we study fractional instantons in Seiberg-Witten theory
\cite{SW}. We will follow the steps of \cite{SW} while relaxing 
the requirement of only integer instanton contributions to the 
low energy effective action.  

Consider $\N=2$ $SU(2)$ Yang-Wills theory in four dimensions. 
The theory is asymptotically 
free in the UV and is strongly coupled below dynamically generated scale 
$\Lambda$. We would like to study the low energy  physics of this theory.  
Classically, the theory has a Coulomb moduli space parameterized by the 
expectation value of the adjoint scalar  
\be
\Phi={\rm diag}(a,-a)\,,
\label{phia}
\ee
in the Cartan subalgebra of the gauge group. 
At the generic point on the moduli space the gauge symmetry is broken to 
$U(1)$. The entire low-energy effective action  $\cL$ of an
Abelian $\N=2$ vector multiplet is completely determined in
terms of the single prepotential $\cF\equiv\cF(\Lambda,a)$ which
depends holomorphically on the strong coupling scale of the 
theory $\L$, and the Coulomb modulus $a$
\bea
8\pi\cL= &&-{\rm Im}[\t]
\left(\partial_\mu a\partial^\mu\bar{a}+i\bar{\psi}\bar{\sigma}^\mu 
\partial_\mu \psi\right) \cr
&& +{\rm Re}\left\{\tau\left(
{i\over 2}F_{\mu\nu}F^{\mu\nu}+
{1\over 2}F_{\mu\nu}\tilde{F}^{\mu\nu}-
2\bar{\lambda} \bar{\sigma}^{\mu} \partial_{\mu}\lambda
\right)\right\}\,,
\label{n2lag}
\eea
with
\be
\tau={d^2 \cF\over d^2 a}\,.
\label{gtau}
\ee
In Eq.~(\ref{n2lag}) $\psi$ and $\lambda$ are fermionic
superpartners of the scalar and the gauge boson respectively.
Classically, the prepotential is given by
\be
\cF_{\rm class}={1\over 2}\tau_0  a^2\,.
\label{fclass}
\ee
where $\t_0={\theta_0\over \pi}+{8\pi i\over g_0^2}$ is the bare
coupling constant. This prepotential receives quantum corrections. 
The tree level and one-loop contributions add up to
\be
\cF_{\rm pert}={i a^2\over \pi} \ln\left[{a^2\over \L^2}\right]\,.
\ee
Higher order perturbative corrections are absent, although 
there are nonperturbative corrections due to instantons.
$\N=2$ lagrangian (\ref{n2lag}) has $U(1)_R$ global 
symmetry\footnote{We assign $U(1)_R$ charge two to $a$.}, which is 
broken by anomaly to a $Z_8$. Thus, 
a full prepotential $\cF$ should at most respect $Z_8$ subgroup
of $U(1)_R$. One instanton action violates $U(1)_R$ symmetry by eight 
units, so {\it assuming} that there are no nonperturbative 
effects that further break this R-symmetry, Seiberg arrived 
at the following form of the full prepotential at weak coupling 
\cite{s88}    
\be
\cF={i a^2\over \pi} \ln\left[{a^2\over \L^2}\right]
+{1\over 2\pi i} a^2 \sum_{\ell=1}^{\infty} c_{\ell} 
\left({\L\over a}\right)^{4\ell}\,,
\label{ffulls}
\ee  
where the $\ell$'th term arises as a contribution of 
$\ell$ instantons. It is this assumption of the exact $Z_8$ 
symmetry of the low energy effective $SU(2)$ 
prepotential that we want to relax. Specifically, 
we assume that in addition to instanton corrections, there are 
nonperturbative corrections which carry $1/2$ unit the 
instanton charge. So we demand only $Z_4$ R-symmetry 
of the quantum prepotential
\be
\cF_{1/2}={i a^2\over \pi} \ln\left[{a^2\over \L^2}\right]
+{1\over 2\pi i} a^2 \sum_{\ell=1}^{\infty} c_{\ell} 
\left({\L\over a}\right)^{2\ell}\,.
\label{ffull}
\ee  

To proceed with the full solution of the model subject to 
(\ref{ffull}), we review physical assumptions of the original 
Seiberg-Witten solution \cite{SW}.  We would like to emphasize 
that in our solution we adopt all constraints listed below.
First, the unitarity constrains ${\rm Im}\{\tau\}>0$ 
throughout the moduli space.
As a result, $\cF$, $a^D\equiv {d \cF\over da}$, $\t$  are defined 
only locally on the moduli space. Low-energy electric-magnetic 
duality \cite{SW} implies that $a$ is a 
multi-valued section on the moduli space as 
well, and thus can not be a nice global coordinate.
Seiberg and Witten thus introduce global coordinate $u$, such that
the period section
\be
\pmatrix{a^D\cr a}\sim \pmatrix{{i\over \pi}\sqrt{2u}\ln\left[{u\over
\Lambda^2}\right]\cr \sqrt{u\over 2}}, \qquad |u|\gg |\Lambda^2|\,.
\label{uinfty}
\ee
The monodromy of the period section due to the semi-classical 
singularity is determined by the asymptotic (\ref{uinfty})  
\be
\pmatrix{a^D\cr a}\to M_{\infty}\pmatrix{a^D\cr a}\equiv
\pmatrix{-1&4\cr 0&-1\cr} \pmatrix{a^D\cr a}\,.
\label{mminfty}
\ee
Second, in addition to the semi-classical singularity on the moduli 
space at $u=\infty$, there are precisely two other singularities 
at $u=\pm \Lambda^2$.
The $u=-\Lambda^2$ singularity is generated by integrating out 
massless dyon of charge $(1,-2)$ (the BPS formula determines its exact mass 
to be $m^{(1,-2)}=|a^D-2 a|$), and the $u=\Lambda^2$ singularity is due to 
the massless monopole of charge $(1,0)$ ($m^{(1,0)}=|a^D|$). 
These two singularities generate the following monodromies of the 
period section  
\bea
&&\pmatrix{a^D\cr a}\to M^{(1,-2)} \pmatrix{a^D\cr a}\equiv
\pmatrix{-1&4\cr -1&3\cr} \pmatrix{a^D\cr a},\ |u+\Lambda^2|\ll 
|\Lambda^2|,
\label{mon2a}
\eea
\bea
&&\pmatrix{a^D\cr a}\to M^{(1,0)} \pmatrix{a^D\cr a}\equiv
\pmatrix{1&0\cr -1&1\cr} \pmatrix{a^D\cr a},\ |u-\Lambda^2|\ll 
|\Lambda^2|.
\label{mon2}
\eea
It is important that (\ref{mon2a}), (\ref{mon2}) 
are determined using low-energy 
electric-magnetic dualities once the charges of the massless 
states on the moduli space are specified\footnote{Actually, one has to specify 
the nature of only one of the two non-perturbative singularities 
on the moduli space. The monodromy (and  charges of a state that generates 
it) due to the other one is determined from the monodromy algebra: 
$M_{\infty}=M^{(1,0)}\cdot M^{(1,-2)}$. }. 
Using above assumptions, Seiberg and Witten identified $\tau(u)$
with the complex structure of the one-parameter family of tori 
\be
y^2=(x^2-u)^2-\Lambda^4\,,
\label{swcurve}
\ee
and the section $(a^D,a)$ with the integral of one-form 
$\lambda\equiv {1\over \sqrt{2}\pi} x^2 {dx\over y}$ over their homology basis.
Note that the strong coupling scale enters as $\Lambda^4$
in (\ref{swcurve}). As a result, in the 
semi-classical region $|a|\gg |\Lambda|$ the prepotential is 
guaranteed to have only integer instanton expansion (\ref{ffulls}).

We now discuss solution of the $SU(2)$ YM theory 
with low energy effective prepotential (\ref{ffull}) 
in the weakly coupled region of the moduli space.   
We put $\L=1$ and assume the existence of a global 
coordinate $f$ on the moduli space. We assume the moduli space 
singularities to be at 
$f=\{0,1,\infty\}$ with the monodromies of the period section
$(a^D(f),a(f))$ given by $\{M^{(1,-2)},M^{(1,0)},M_{\infty}\}$
respectively. 
The weak coupling asymptotic is assumed to be 
\be
\pmatrix{a^D\cr a}\sim\pmatrix{{2i\over \pi}\ \sqrt{f}\ \ln f\cr \sqrt{f}},\qquad 
f\gg 1\,.
\label{wwf}
\ee
Comparing (\ref{wwf}) with (\ref{uinfty}) we thus have 
\be
f\sim {u\over 2\Lambda^2},\qquad |u|\gg |\Lambda^2|\,.
\label{fu}
\ee
The construction of $SL(2,Z)$ sections $(a^D(f),a(f))$ with 
required weak coupling asymptotic (\ref{wwf}) and 
monodromies (\ref{mminfty}), (\ref{mon2}) and (\ref{mon2a}) is rather simple.
We start with the following ansatz for $a^D$:
\be
a^D=A \left(1-{1\over f}\right)^{\d_1+1} f^{\d_2+1/2}\ 
_2F_1\left(\a,\b,\a+\b+m,1-{1\over f}\right)\,,
\label{ad1}
\ee
where $A$ is a normalization constant and  $m$ is an integer. 
Above ansatz insures that the only singularities of $a^D$ occur 
at $f=\{0,1,\infty\}$. The third parameter 
in the hypergeometric function,  ($\a+\b+m$), is chosen to
get a logarithmic singularity in $a^D$ as $f\to\infty$.   
Furthermore, we assume that $\G(\a+\b+m)$ is finite to 
have (\ref{ad1}) well-defined.
A set of useful identities among  hypergeometric functions 
can be found in \cite{htf}. 
Comparing the asymptotics of  (\ref{ad1}) as $f\to \infty$ 
with (\ref{wwf}) we find
\bea
m&=&0\,,\cr
\d_2&=&0\,,\cr
A&=&{2i\over \pi}\ {\G(\a) \G(\b)\over \G(\a+\b)}\,.
\label{data}
\eea
The monodromy of (\ref{ad1}) 
around $f=\infty$ determines $a$. With (\ref{data}), 
we find 
\be
a=f^{1/2} \left(1-{1\over f} \right)^{\d_1+1}\ _2F_1
\left(\a,\b,1,{1\over f}\right)\,. 
\label{a}
\ee  
Using identifies of \cite{htf} 
it is straightforward to check that the monodromy of 
$(a^D,a)$ about $f=1$ requires 
 \bea
\a&=&{1\over 2}+n\,,\cr
\b&=&{1\over 2}-n+k\,,\cr
\delta_1&=&d\,,
\eea 
where $n,d$ are arbitrary integers, and  $k$ is a non-negative 
integer.
Altogether, $SL(2,Z)$
sections $S(k,n,d)$ 
\bea
&&\S(k,n,d)\equiv\pmatrix{a^D\cr\cr a\cr}\cr
&=&\pmatrix{
{2i\over \pi}\ {\G({1\over 2}+n) \G({1\over 2}-n+k)\over \G(k+1)}\ 
f^{1/2} \left(1-{1\over f} \right)^{d+1}\ _2F_1
\left({1\over 2}+n,{1\over 2}-n+k,k+1,1-{1\over f}\right)\cr
\cr
f^{1/2} \left(1-{1\over f} \right)^{d+1}\ _2F_1
\left({1\over 2}+n,{1\over 2}-n+k,1,{1\over f}\right)\cr}\cr
&&\ 
\label{fS}
\eea
parameterized by integers $(n,k,d)$ with $k\ge 0$, satisfy 
monodromies (\ref{mon2}), (\ref{mminfty}) about $f=\{1,\infty\}$
punctures of the $f$-sphere. As the only other singularity of 
$S(k,n,d)$ is at $f=0$, the monodromy (\ref{mon2a}) is satisfied 
automatically. 

It is easy to see that if (\ref{fS}) solves the monodromy problem, 
the metric on the moduli space is positive definite. Really, 
since the structure group of the period section is a subgroup of $SL(2,Z)$,
the effective coupling 
\be
\tau_{(k,n,d)}(f)= {{d a^D/ d f} \over {d a/ d f} }\,,
\label{tauf}
\ee
is a section with the same structure group and thus 
${\rm Im}{\tau_{(k,n,d)}}$ can not 
change sign over the whole moduli space. The original Seiberg-Witten 
solution \cite{SW} 
is realized by section $\S(1,0,0)$ (or equivalently $\S(1,1,0)$).
Really, substituting $f \equiv(u+\Lambda^2)/2\Lambda^2$ in (\ref{fS}) 
and  using certain identities for the hypergeometric functions \cite{htf},
we recover 
\be
\S(1,0,0)
=\pmatrix{
{i u^{1/2}\over 4}\ \left(1-{\L^4\over u^2} \right)\ _2F_1
\left({3\over 4},{5\over 4},2,1-{\L^4\over u^2}\right)\cr
\cr
{u^{1/2}\over \sqrt{2}} \ _2F_1
\left(-{1\over 4},{1\over 4},1,{\L^4\over u^2}\right)\cr}\,. 
\label{orSW}
\ee

A straightforward computation shows that the effective coupling 
$\tau_{(k,n,d)}$ has generically fractional instanton corrections in 
the semi-classical region of the moduli space $|a|\gg 1$,
\be
\t_{(k,n,d)}(a)={2i\over \pi}\biggl[\ln a^2+t_0+{t_1\over a^2}
+{t_2\over a^4}+{t_3\over a^6}+\cdots+{t_p\over a^{2p}}+\cdots\biggr]\,,
\label{asstau}
\ee
with 
\bea
t_0&=&2+2\psi(1)-\psi\left({1\over 2}+n\right)-
\psi\left({1\over 2}-n+k\right)\,,\cr
\cr
t_1&=&k-1-2d\,,\cr
\cr
t_2&=&12\,d^2 - 12\,d\,\left( -1 + k + k\,n - n^2 \right)
+\biggl[3\,\biggl( 59 + 136\,n^2 + 48\,n^4 \cr 
&&+
        4\,k^2\,\left( 17 + 32\,n + 12\,n^2 \right)  -
        8\,k\,\left( 16 + 17\,n + 16\,n^2 + 12\,n^3 \right)  \biggr) 
\biggr]/64\,,\cr
&&
\label{ts}
\eea
where 
\be
\psi(z)\equiv {d\ln\G(z)\over d z}\,.
\label{psidef}
\ee

Though any section $\S(k,n,d)$  with $k\ge 0$ solves the Seiberg-Witten 
monodromy problem, restrictions on $(k,n,d)$ come 
from the conjectured spectrum of BPS states at the singularities. 
Suppressing numerical constants, we have    
\bea
a^D\sim a&\sim& f^{k-n-d}\ln f+f^{n-d}\cr
a^D-2a&\sim& f^{k-n-d},\qquad\qquad\qquad {\rm if}\ k>2n,\cr
\cr
a^D\sim a&\sim& f^{k/2-d}\ln f\cr
a^D-2a&\sim& f^{k/2-d},\qquad\qquad\qquad {\rm if}\ k=2n,\cr
\cr
a^D\sim a&\sim& f^{n-d}\ln f+f^{k-n-d}\cr
a^D-2a&\sim& f^{n-d},\qquad\qquad\qquad {\rm if}\ k<2n\,,
\label{aa0}
\eea
as $f\to 0$, and 
\bea
a^D&\sim& (f-1)^{d+1}\cr
a&\sim&(f-1)^{d+1-k}+(f-1)^{d+1}\ln(f-1),\qquad\qquad\qquad {\rm if}\ 
k>0,\cr
\cr
a^D&\sim& (f-1)^{d+1}\cr
a&\sim&(f-1)^{d+1}\ln(f-1),\qquad\qquad\qquad {\rm if}\ k=0\,,
\label{aa1}
\eea
as $f\to 1$. Thus, from (\ref{aa1}), to have a massless monopole 
at $f=1$ and massive all the electrically charged particles
we must have
\be
k\ge d+1\ge 1\,.
\label{con1}
\ee 
Similarly, from (\ref{aa0}), 
to have {\it only} massless dyon of charge $(1,-2)$ at 
$f=0$ singularity 
\be
\biggl\{n\le {\rm min}[d, k-d-1]\biggr\}\ \bigcup\ 
\biggl\{n\ge {\rm max}[d+1,k-d]\biggr\}\,.
\label{con0}
\ee
Constraints  (\ref{con1}), (\ref{con0}) are mutually 
compatible;  thus is appears we found new solutions 
to the $\N=2$ $SU(2)$ monodromy problem with the 
same weak coupling asymptotic and the same massless 
states at nonperturbative singularities on the 
moduli space as in \cite{SW}. Generically, new solutions have 
$1/2$-instanton corrections in the semi-classical 
region of the moduli space. 

In the rest of this section we show that all new  solutions 
(note that $\S(1,0,0)\equiv \S(1,1,0)$  corresponds to 
the Seiberg-Witten solution) are in fact unphysical: they 
predict that giving mass to the chiral multiplet $\Phi$ 
in $\N=2$ vector multiplet breaks the supersymmetry completely.     
On the contrary, when a mass of $\Phi$ is much larger than the strong
coupling scale of the $\N=2$ theory, we should be able to 
reliably integrate it out, thus ending up with $\N=1$ $SU(2)$ 
Yang-Mills theory which is predicted to have two vacua \cite{w82}.  
The analysis below repeat those of \cite{SW}. 

Breaking 
$\N=2$ supersymmetry down to $\N=1$ is achieved by adding
a superpotential for the chiral multiplet in $\N=2$ vector 
multiplet
\be
W=m {\rm Tr} \Phi^2\,.
\label{mass}
\ee 
In the low energy effective theory the operator ${\rm Tr} \Phi^2$
is represented by a chiral superfield $f$. Its first component is the 
scalar field $f$ whose expectation value is 
\be
<f>\ =\ <{\rm Tr}\Phi^2>\,.
\label{vevf}
\ee 
It was argued in \cite{SW} that adding (\ref{mass}) microscopically
corresponds to adding 
\be
W_{eff}=m f\,,
\label{masse}
\ee
to the low energy effective superpotential. 
At a generic point on the moduli space there are no light 
chiral fields, so (\ref{masse}) is the complete superpotential. 
Thus perturbation (\ref{masse}) lifts all such $\N=2$ vacua.
The situation is different near the singularities on the moduli 
space. Near the $f=1$ singularity there are massless monopoles.
The monopoles can be represented by ordinary (local) chiral 
superfields $M$ and $\tilde{M}$, as long as we describe the 
gauge field by the dual to the original photon, $a^D$. The 
complete superpotential is then 
\be
W_{f=1}=\sqrt{2}a^D M \tilde{M}+m f\,,
\label{mass1}
\ee
where the first term represents $\N=2$ superpotential of the $m=0$
theory. F-term equations from (\ref{mass1}) give
\bea
\sqrt{2} M \tilde{M} +m {df \over da^D}&=&0\,,\cr
\cr
a^D M=a^D \tilde{M}&=&0\,.
\label{F1}
\eea
Using (\ref{aa1}), eq.~(\ref{F1}) has a solution (there is $\N=1$ vacuum) 
provided  
\be
{df \over da^D}\ne \infty\qquad {\rm  at}\ f=1\,.   
\label{cond1}
\ee
Along with (\ref{con1}), eq.~(\ref{cond1}) implies that 
\be
d=0,\ k\ge 1\,.
\label{d0}
\ee
Identical analysis at the dyon singularity, $f=0$, shows that 
$\N=1$ vacuum there exists provided
\be
{df \over d(a^D-2a)}\ne \infty\qquad {\rm  at}\ f=0\,.
\label{cond0}
\ee
Eqs.~(\ref{con0}) and (\ref{cond0}) give
\be
\biggl\{n=k-d-1,\ 2d+1\ge k\biggr\}\ \bigcup\ \biggl\{n=d+1,\ 
2d+1\ge k\biggr\}\,.
\label{nk}
\ee  
Combining (\ref{d0}) and (\ref{nk}) we conclude 
that only sections $\S(1,0,0)\equiv\S(1,1,0)$ predict a pair 
of $\N=1$ supersymmetric vacua for the mass deformed 
$\N=2$ $SU(2)$ YM theory. These sections are precisely 
the Seiberg-Witten solution of the model, which do not 
have fractional instantons in the semi-classical region of the 
moduli space.

\section{Conclusion}
D3 brane probe computation in  gravitational dual of large-N gauge theories 
with $\N=2$ supersymmetry suggests that, unlike $\N=1$  supersymmetric 
gauge theories, these theories 
do not have fractional instantons. The evidence comes primarily from the 
facts that the enhancon is a sharp boundary of a D-brane
probe moduli space, 
and the agreement of the metric on the probe moduli 
space with the one-loop beta-function computation in the 
dual gauge theory. 

In this paper we studied fractional instantons in the 
$\N=2$ supersymmetric gauge theories from the field 
theoretical perspective. On the example of $SU(2)$
Yang-Mills theory we showed that though it is 
possible to construct new solutions to the Seiberg-Witten 
monodromy problem with the same perturbative asymptotic,
but fractional instanton corrections in the semi-classical 
region of the moduli space, these  solutions are unphysical.
Specifically, they predict that the soft mass term to the 
chiral superfield in $\N=2$  vector multiplet 
breaks the supersymmetry completely. 
Our analysis points out that allowing fractional 
instantons in the semi-classical prepotential would 
drive monopole (dyon) condensate to infinity 
in mass deformed $\N=2$ theories.

\section*{Acknowledgements}

I would like to thank Per Berglund, Clifford Johnson, Amanda Peet  and
Joe Polchinski for valuable discussions.  
This work was supported in part by NSF grants PHY97-22022 and PHY99-07949.



\newpage
\begin{thebibliography}{99}

\bibitem{gsr} M.~T.~Grisaru, W.~Siegel and M.~Rocek,
{\it Nucl.\ Phys.}\  {\bf B159} (1979) 429. 

\bibitem{pertprep}
V.~A.~Novikov, M.~A.~Shifman, A.~I.~Vainshtein, M.~B.~Voloshin and
V.~I.~Zakharov, {\it Nucl.\ Phys.}\ {\bf B229} (1983) 394; V.~A.~Novikov,
M.~A.~Shifman, A.~I.~Vainshtein and V.~I.~Zakharov, {\it Nucl. Phys.}\ 
{\bf B229} (1983) 381, 407; V.~A.~Novikov, M.~A.~Shifman, A.~I.~Vainshtein and
V.~I.~Zakharov, {\it Phys. Lett.}\ {\bf B166} (1986) 329.

\bibitem{n4} M.~F.~Sohnius and P.~C.~West, 
{\it Phys. Lett.}\ {\bf B100} (1981) 245; S.~Mandelstam, 
{\it Nucl.\ Phys.}\  {\bf B213} (1983) 149.


\bibitem{s88} N.~Seiberg, 
{\it Phys. Lett.}\ {\bf B206} (1988) 75.


\bibitem{kss9706} A.~Kovner, M.~Shifman and  A.~Smilga,
{\it Phys. Rev.}\ {\bf D56} (1997) 7978, hep-th/9706089.

\bibitem{b9803} J.~Brodie, {\it Nucl.Phys.}\  {\bf B532} (1998) 137,
hep-th/9803140.

\bibitem{dk0001} N.~Dorey and S.~P.~Kumar, {\it JHEP} {\bf 0002} 
(2000) 006, hep-th/0001103.


\bibitem{ps0003} J.~Polchinski and M.~J.~Strassler,
hep-th/0003136.

\bibitem{w97} E.~Witten, {\it Nucl.Phys.}\ {\bf B500} (1997) 3,
hep-th/9703166.

\bibitem{bpp0008} A.~Buchel, A.~W.~Peet and J.~Polchinski,
hep-th/0008076.


\bibitem{ejp} N.~Evans, C.~V.~Johnson and M.~Petrini,
{\it JHEP } {\bf  0010} (2000) 022,
hep-th/0008081.



\bibitem{SW} 
N.~Seiberg and E.~Witten,
{\it Nucl.\ Phys.}\  {\bf B426} (1994) 19, hep-th/9407087.


\bibitem{w82} E.~Witten, {\it Nucl.\ Phys.}\  {\bf B202} (1982) 253.



\bibitem{juan}
J.~Maldacena,
{\it Adv.\ Theor.\ Math.\ Phys.\ }  {\bf 2} (1998) 231,
hep-th/9711200.


\bibitem{klt} P.~Kraus, F.~Larsen and  S.~P.~Trivedi,
{\it JHEP}\ {\bf 9903} (1999) 003, hep-th/9811120.

\bibitem{pw}  K.~Pilch and N.~P.~Warner, 
hep-th/0004063.



\bibitem{dw}
R.~Donagi and  E.~Witten, {\it Nucl.Phys.}\ {\bf B460} (1996) 299,
hep-th/9510101.


\bibitem{ds} M.~R.~Douglas and S.~H.~Shenker,
{\it Nucl.Phys.}\ {\bf B447} (1995) 271, hep-th/9503163.

\bibitem{jpp9911} C.V.~Johnson, A.W.~Peet and J.~Polchinski,
hep-th/9911161.

\bibitem{htf}  H.~Bateman, ``Higher Transcendental Functions'',
Volume I, McGraw-Hill Book Company, 1953. 


\end{thebibliography}
\end{document}